\documentclass[a4paper,conference]{IEEEtran}

\usepackage{amsfonts}
\usepackage{mathtools}
\usepackage{wasysym}

\usepackage{amssymb}
\usepackage{amsmath}
\usepackage{bbm}
\usepackage{bm}
\usepackage{dsfont}
\usepackage{graphicx}

\usepackage{enumitem}
\usepackage{tikz}
\usepackage{subfigure}
\usepackage{ifpdf}

\newcommand{\dH}{d_{\text{H}}}
\newcommand{\wH}{w_{\text{H}}}

\newcommand{\Eex}{E_{\text{ex}}}
\newcommand{\Ex}{E_{\text{x}}}
\newcommand{\Pe}{\mathsf{P}_{\text{e}}}

\newcommand{\Pemax}{\mathsf{P}_{\text{e,max}}}

\newtheorem{theorem}{Theorem}
\newtheorem{proposition}{Proposition}

\newtheorem{remark}{Remark}
\newtheorem{corollary}{Corollary}

\newcommand{\matc}{\ensuremath{\mathcal{C}}}

\ifx\eqref\undefined
	\newcommand{\eqref}[1]{~(\ref{#1})}
\fi
\ifx\mod\undefined
	\def\mod{\mathop{\rm mod}}
\fi

\def\eqdef{\stackrel{\triangle}{=}}

\ifx\lesssim\undefined
\def\simleq{{{\mskip 3mu plus 2mu minus 1mu%
	\setbox0=\hbox{$\mathchar"013C$}%
	\raise.2ex\copy0\kern-\wd0%
	\lower0.9ex\hbox{$\mathchar"0218$}}\mskip 3mu plus 2mu minus 1mu}}
\else
\def\simleq{\lesssim}
\fi
\ifx\gtrsim\undefined
\def\simgeq{{{\mskip 3mu plus 2mu minus 1mu%
	\setbox0=\hbox{$\mathchar"013E$}%
	\raise.2ex\copy0\kern-\wd0%
	\lower0.9ex\hbox{$\mathchar"0218$}}\mskip 3mu plus 2mu minus 1mu}}
\else
\def\simgeq{\gtrsim}
\fi

\begin{document}

%\sloppy
\IEEEoverridecommandlockouts

\thinmuskip=2mu plus 1mu minus 3mu
\medmuskip=2mu plus 2mu minus 1mu
\thickmuskip=2mu plus 2mu minus 2mu

%% Paper Title
%% You can use linebreaks \\ within to get better formatting as
%% desired. 
\title{Bounds on the Reliability of a Typewriter Channel} 

\author{
\IEEEauthorblockN{Marco Dalai}
\IEEEauthorblockA{University of Brescia\\
marco.dalai@unibs.it}
\and
\IEEEauthorblockN{Yury Polyanskiy}
\IEEEauthorblockA{Massachusetts Institute of Technology\\
yp@mit.edu}%
\thanks{
The research was supported by the NSF grant CCF-13-18620 and NSF Center for Science of Information (CSoI) 
under grant agreement CCF-09-39370. The work was mainly completed while the authors were visiting the Simons Institute for the Theory of Computing at UC Berkeley, whose support is gratefully acknowledged. }
}

\maketitle

\begin{abstract}
We give new bounds on the reliability function of a typewriter channel with 5 inputs and crossover probability $1/2$. 
The lower bound is more of theoretical than practical importance; it improves very marginally the expurgated bound, providing a counterexample to a conjecture on its tightness by Shannon, Gallager and Berlekamp which does not need the construction of algebraic-geometric codes previously used by Katsman, Tsfasman and Vl\u{a}du\c{t}.
The upper bound is derived by using an adaptation of the linear programming bound and it is essentially useful as a low-rate anchor for the straight line bound. 

\end{abstract}

\section{Introduction}

Consider the typewriter channel $W:\mathbb{Z}_5\to \mathbb{Z}_5$ with five inputs \cite[Fig. 2]{shannon-1956} and crossover probability $1/2$.
%defined by
%\begin{equation}
%W(y|x)=
%\begin{cases}
%1/2 & y=x \mbox{ or } y=x+1\\
%0 & \mbox{otherwise}
%\end{cases}
%\end{equation}
%We consider the memoryless extension of the channel $\bm{W}$ such that for $\bm{x}=(x_1,\ldots,x_n)$ and $\bm{y}=(y_1,\ldots,y_n)$
%\begin{equation}
%\bm{W}(\bm{y}|\bm{x})=\prod_{k=1}^n W(y_i|x_i).
%\end{equation}
%Define the reliability function of the channel as
%\begin{equation}
%E(R)=\limsup_{n\to\infty} \frac{1}{n}\log \Pe(\lceil e^{n R} \rceil, n)
%\end{equation}
%where $\Pe(M, n)$  is the smallest possible probability of error\footnote{It is well known that $\Pe$ can be taken to be the average probability of error or maximum probability of error over codewords, without affecting the definition of $E(R)$.} of codes with $M$ codewords of length $n$.
This channel has a great historical importance \cite{shannon-1956}, \cite{lovasz-1979}. In this paper we study the problem of bounding its reliability function $E(R)$ defined by
\begin{equation*}
E(R)=\limsup_{n\to\infty} \frac{1}{n}\log \frac{1}{\Pe(\lceil 2^{n R} \rceil, n)}
\end{equation*}
where $\Pe(M, n)$  is the smallest possible probability of error\footnote{In particular, the definition of $E(R)$ does not depend on whether we use maximal or average probability of error over codewords, see \cite{shannon-gallager-berlekamp-1967-1}.} of codes with $M$ codewords of length $n$. Here and below, rates are in bits and all logarithms are taken to base 2.

The interval of interest is $C_0<R< C$, where $C_0=\log\sqrt{5}$ is the zero-error capacity and $C= \log(5/2)$ is the ordinary capacity, since $E(R)=+\infty$ for $R\leq C_0$ and $E(R)=0$ for $R\geq C$. All equations below should be interpreted as restricted to this interval. To the best of our knowledge, the best known bounds in the literature date back to \cite{gallager-1965}, \cite{shannon-gallager-berlekamp-1967-2}, \cite{lovasz-1979} and reduce to the following (see Section \ref{sec:previous_bounds} for a detailed discussion of these bounds).
%\begin{itemize}
%\item
\begin{proposition}[Random/expurgated bound \cite{gallager-1965}] 
\label{prop:expurgated}
We have $E(R)\geq E_{\text{r/ex}}(R)$ where
\begin{equation}
E_{\text{r/ex}}(R)= 
\log(5/2)-R\,.
\end{equation}
\end{proposition}

\begin{proposition}[Straight line bound \cite{shannon-gallager-berlekamp-1967-2}]
\label{prop:straight_line}
We have $E(R)\leq E_{\text{sl}}(R)$ where
\begin{equation}
E_{\text{sl}}(R)=
({\log(\sqrt{5}/2)})^{-1} (\log(5/2)-R)\,.
\end{equation}
\end{proposition}
%\item 
%\end{itemize}

\begin{figure}
\includegraphics[width=\linewidth]{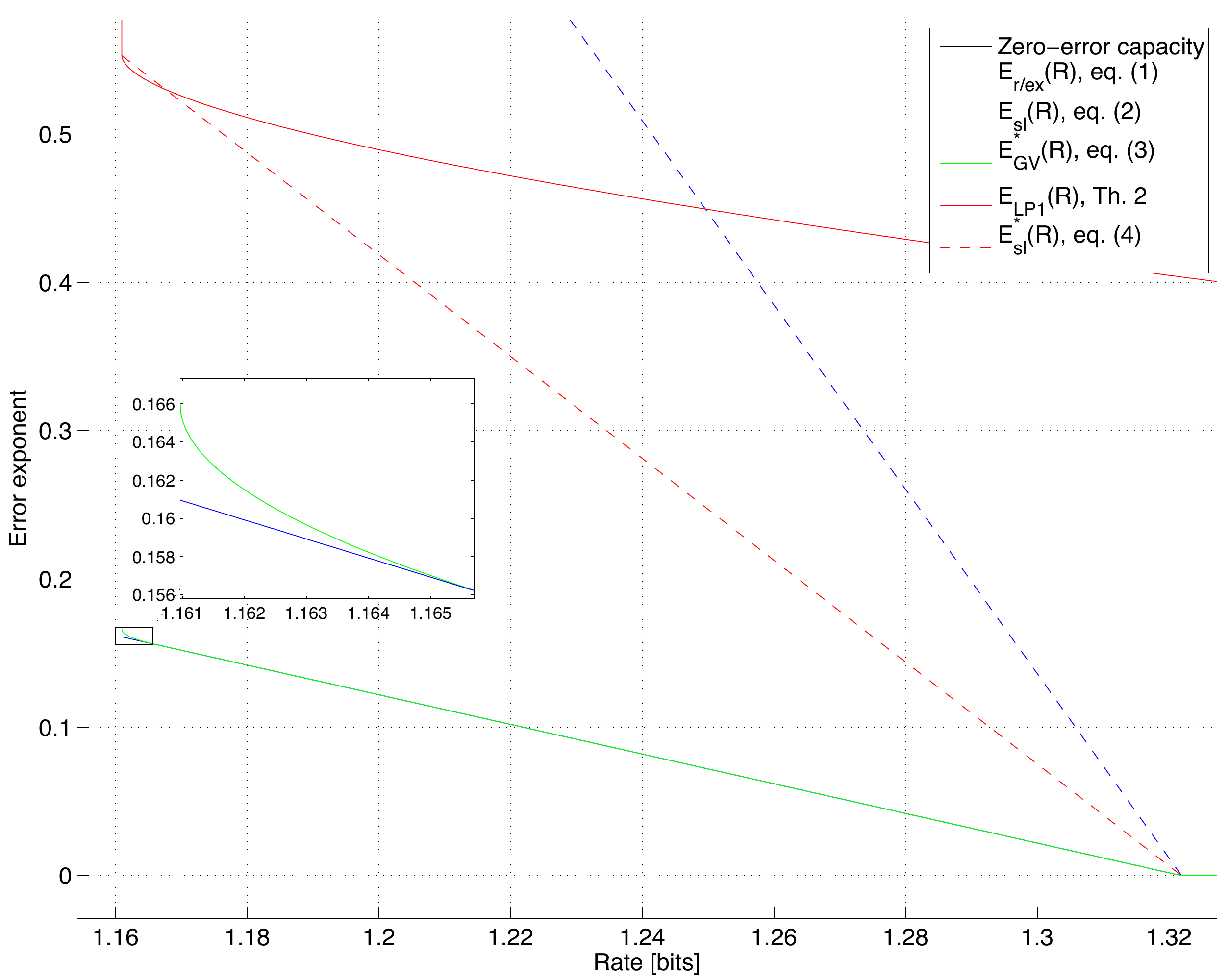}
\caption{Bounds for a $5$-input typewriter channel with cross-over probability $1/2$.}
\label{fig:typewriter-5}
\end{figure}

Let $H_q(t)$ be the $q$-ary entropy function defined as
\begin{equation*}
H_q(t) = t\log(q-1)-t\log t - (1-t)\log(1-t)\,.
\end{equation*}
In this paper we prove the following bounds (see Figure \ref{fig:typewriter-5}):

\begin{theorem}
\label{th:GV-lower}
We have $E(R)\geq E_{\text{GV}}^*(R)$ where
\begin{align}
E_{\text{GV}}^*(R) & =\begin{cases}
\left(\frac{4}{3}-H_2(1/3)\right)\delta(R) &\log\sqrt{5}\leq R\leq R^*\\
E_{\text{r/ex}}(R) &\mbox{ otherwise }
\end{cases}\,,\\
R^* & =\log(5)-\frac{1}{2}H_2(1/4)-\frac{3}{4}\,,\nonumber
\end{align}
and $\delta(R)$ is the solution of the equation
\begin{equation*}
R=\log(5)-2\delta  + \frac{1}{2}H_2(2\delta).
\end{equation*}
\end{theorem}

%\begin{theorem}
%\label{th:ELP}
%We have $E(R)\leq E_{\text{LP}}(R)$ where $E_{\text{LP}}(\cdot)$ is the inverse function of the function $R_{LP'}(\cdot)$ defined by
%\begin{align}
%R_{LP'}(E)& = \frac{1}{2}\log(5) + H_q\left({(q-1)-(q-2)\delta - 2\sqrt{(q-1)\delta(1-\delta)}\over q}\right)\,,\label{eq:R_LP}\\
%   H_q(x) & = x \log(q-1) - x \log x - (1-x)\log(1-x)\,,\\
%   \delta & = E/\log(2)
%\end{align}
%and, hence, $E(R)\leq E_{sl,LP}(R)$ where $E_{sl,LP}(R)$ is the tangent drawn from $E_{\text{LP}}(R)$ to the point $(\log(5/2),0)$ in the $(R,E)$ plane.
%\end{theorem}

\begin{theorem}
\label{th:ELP}
We have $E(R)\leq E_{\text{LP1}}(R)$ where $E_{\text{LP1}}(\cdot)$ is defined implicitly by its inverse function 
\begin{equation*}
E_{\text{LP1}}^{-1}(E)=\log\sqrt{5}+R_{\text{LP1}}\left(\sqrt{5},E\right)
\end{equation*}
where
\begin{align*}
R_{\text{LP1}}(q,\delta) &= H_q\left({(q-1)-(q-2)\delta - 2\sqrt{(q-1)\delta(1-\delta)}\over q}\right)
\end{align*}
is the linear programming bound for codes in a $q$-ary Hamming space.
\end{theorem}

Since $E_{\text{LP1}}(\log \sqrt{5})=(1-1/\sqrt{5})$, we obtain the following improvement of the straight line bound\footnote{Anchoring the straight line bound to $E_{\text{LP1}}$ at $R=\log\sqrt{5}$ is only very marginally suboptimal, as is seen from Figure \ref{fig:typewriter-5}.}.
\begin{corollary}
$E(R)\leq E_{\text{sl}}^*(R)$ where
\begin{equation}
E_{\text{sl}}^*(R)=
\left(1-\frac{1}{\sqrt{5}}\right)E_{\text{sl}}(R)\,.
\label{eq:star-straight-line}
\end{equation}
\end{corollary}

\begin{remark}
While Theorem \ref{th:GV-lower} is expecially derived for the channel with five inputs, Theorem \ref{th:ELP} can be extended to any odd number of inputs, and the proof in Section \ref{sec:converse-proof} is given in this more general form.
\end{remark}

\section{Discussion of Propositions 1 and 2}
\label{sec:previous_bounds}

In this section we prove that the best previously known bounds on $E(R)$ are those given by Propositions \ref{prop:expurgated} and \ref{prop:straight_line}.

We first prove that the bound given in Proposition \ref{prop:expurgated} corresponds to the best possible one which can be derived from Gallager's random coding and expurgated bounds \cite{gallager-1965}, even when computed on blocks of arbitrary lengths. For this particular channel, since the capacity equals the cutoff rate, it suffices to consider the expurgated bound. 

Let $g_n:\mathbb{Z}_5^n\times\mathbb{Z}_5^n\to\mathbb{R}$ be the function defined by
\begin{equation*}
g_n(\bm{x}_1,\bm{x}_2)=\sum_{\bm{y}\in \mathbb{Z}_5^n}\sqrt{\bm{W}(\bm{y}|\bm{x}_1)\bm{W}(\bm{y}|\bm{x}_2)},
\end{equation*}
where $\bm{W}$ is the $n$-fold memoryless extension of $W$, and define
\begin{align*}
%\Ex^n(\rho,P^n) & =-\frac{\rho}{n}\log \sum_{\bm{x}_1,\bm{x}_2}
%P^n(\bm{x}_1)P^n(\bm{x}_2)g_n(\bm{x}_1,\bm{x}_2)^{1/\rho}\,,\\
Q^n(\rho,P^n) & =\sum_{\bm{x}_1,\bm{x}_2}
 P^n(\bm{x}_1)P^n(\bm{x}_2)g_n(\bm{x}_1,\bm{x}_2)^{1/\rho}\,,\\ 
\Ex^n(\rho) & =-\frac{\rho}{n}\log\min_{P^n}Q^n(\rho,P^n)\,.
\end{align*}
Gallager's bound \cite{gallager-1965} can be stated as $E(R)\geq \Eex^n(R)$, where
\begin{equation}
\Eex^n(R)= \sup_{\rho\geq 1} \left[\Ex^n(\rho)-\rho R\right].
\label{eq:def-Eex^n}
\end{equation}
This bound holds for any $n$ and hence it makes sense to consider the  optimal bound $\sup_n \Eex^n(R)$. Since the function $n\Eex^n(R)$ is super-additive in $n$%
%\footnote{Consider, in the optimization of $P^{n+m}$, using $P^{n+m}=P^n\times P^m$ where $P^n$ and $P^m$ are optimal for lengths $n$ and $m$ respectively, and proceed as in \cite{gallager-book}.}
, by Fekete's lemma we have
\begin{equation*}
\sup_n \Eex^n(R) = \lim_{n\to \infty} \Eex^n(R).
\end{equation*}
We thus focus on the limit $\Eex^\infty(R)$, which gives the best bound $E(R)\geq \Eex^\infty(R)$.
Computing $\Eex^n(R)$ for a general channel is prohibitive even for small values of $n$. However, for the considered typewriter channel, we can determine
%	\footnote{We observe that even if we give a closed form expression for $\Eex^2(R)$ and $\Eex^\infty(R)$, we do not know a closed form expression for $\Eex^n(R)$ for odd $n$. In particular, as far as we know, even determining the point at which $\Eex^n(R)$ diverges, which is the highest rate of zero-error codes in the odd powers of the pentagon, is an unsolved problem.}
 $\Eex^\infty(R)$.  To the best of our knowledge the following proposition has not been reported before in the literature.
\begin{proposition}
\label{prop:ex_infty}
For the considered channel, we have  $\Eex^\infty(R)=\Eex^2(R)=E_{\text{r/ex}}(R)$.
\end{proposition}

\begin{IEEEproof}
Consider first the minimization of the quadratic form $Q^n(\rho,P^n)$
and note that the $5^n\times 5^n$ matrix with elements $g_n(\bm{x}_1,\bm{x}_2)^{1/\rho}$, call it $g_n^{\odot \frac{1}{\rho}}$,  is the $n$-fold Kronecker power of the $5\times 5$ matrix
\begin{equation*}
g_1^{\odot \frac{1}{\rho}}=\left(
\begin{array}{ccccc}
1 & \alpha & 0 & 0 & \alpha\\
\alpha & 1 & \alpha & 0 & 0\\
0 & \alpha & 1 & \alpha & 0\\
0 & 0 & \alpha & 1 & \alpha\\
\alpha & 0 & 0 & \alpha & 1
\end{array}
\right),\quad \alpha=2^{-\frac{1}{\rho}}\,.
\end{equation*}
As observed by Jelinek \cite{jelinek-1968}, if $g_1^{\odot \frac{1}{\rho}}$ is a positive semidefinite matrix, so is $g_n^{\odot \frac{1}{\rho}}$. In that case, the quadratic form defining $Q^n(\rho,P^n)$ for any $n$ is a convex function of $P^n$. Jelinek showed that it is minimized by an i.i.d. distribution $P^n=P\times P \cdots\times P$, where $P$ is optimal for $n=1$, and the achieved minimum is just the $n$-th power of the minimum achieved for $n=1$. Thus, if the matrix with elements $g_1(x_1,x_2)^{1/\rho}$ is positive semidefinite, then $\Ex^n(\rho)=\Ex^1(\rho)$. Furthermore, in this case the convexity of the quadratic form and the symmetry of $g_1^{\odot \frac{1}{\rho}}$ imply that the uniform distribution is optimal. Hence, by direct computation,
\begin{equation}
\Ex^n(\rho)=-\rho\log \left(\frac{1}{5}+\frac{2^{1-1/\rho}}{5}\right)
\label{eq:Ex_pos_semidef}
\end{equation}
whenever $g_1^{\odot \frac{1}{\rho}}$ is positive semidefinite. Since its eigenvalues are 
$\lambda_k=1+2^{1-1/\rho}\cos(2\pi k/5)$, $k=0,\ldots, 4$, the matrix is positive semidefinite for $\rho\leq \bar{\rho}= \log 2/\log (2\cos(\pi/5)) \approx 1.4404$.

For $\rho>\bar{\rho}$, $g_1^{\odot \frac{1}{\rho}}$ is not positive semidefinite, and the minimization of $Q^n(\rho,P^n)$ is in general difficult to study. We prove that $\Ex^n(\rho)\leq  \rho\log(5)/2$ and then show that $\Ex^2(\rho)\geq  \rho\log(5)/2$, which implies $\Ex^\infty(\rho)=\Ex^2(\rho)=\rho\log(5)/2$. To prove this, observe that the minimum of $Q^n(\rho,P^n)$ is non-decreasing in $\rho$, and hence for $\rho>\bar{\rho}$
\begin{align*}
\Ex^n(\rho) & \leq -\frac{\rho}{n}\log \min_{P^n}Q^n(\bar{\rho},P^n)\\
& = \frac{\rho}{\bar{\rho}}\Ex^n(\bar{\rho})\\
& = \rho \log (5)/2\,,
\end{align*}
where the last step is obtained using the definition of $\bar{\rho}$ and equation \eqref{eq:Ex_pos_semidef}.
To prove that $\Ex^2(\rho)\geq  \rho\log(5)/2$, simply evaluate the function $Q^2(\rho,P^n)$ when $P^2$ is the indicator function of Shanon's zero-error code of length two for the pentagon.

So, we have finally proved that
\begin{equation*}
\Ex^\infty(\rho)=\Ex^2(\rho)=
\begin{cases}
-\rho\log \left(\frac{1}{5}+\frac{2^{1-1/\rho}}{5}\right) &\mbox{ if } \rho\leq \bar{\rho}\\
\rho \log(5)/2 &\mbox{ if } \rho>\bar{\rho}\\
\end{cases}\,.
\end{equation*}
Explicit computation of $\Eex^2(R)$ then reveals that the supremum over $\rho\geq 1$ is achieved by $\rho=1$ if $R\geq \log(5)/2$ and as $\rho\to\infty$ if $R<\log(5)/2$, proving $\Eex^2(R)=E_{\text{r/ex}}(R)$. 
\end{IEEEproof}

Note in particular that, for $R\geq \log(5)/2$, $\Eex^2(R)$ coincides with the simple random coding bound \cite{gallager-1965}, while it is infinite for $R<\log(5)/2$ as implied by the known existence of zero-error codes at those rates \cite{shannon-1956}.

Proposition \ref{prop:ex_infty} combined with \cite{gallager-1965} implies the proof of Proposition \ref{prop:expurgated}. Since, to the best of out knowledge, no previous improvement of the bound $\Eex^\infty(R)$ was known for this channel, it also implies that $E_{\text{r/ex}}(R)$ is the best bound on $E(R)$ deducible from the known results in the literature.

\begin{remark}
The quantity $\Eex^\infty(R)$ was conjectured\footnote{Citing from \cite{shannon-gallager-berlekamp-1967-1}: ``The authors would all tend to conjecture [...] As yet there is little concrete evidence for this conjecture.''} to equal the true reliability function in \cite{shannon-gallager-berlekamp-1967-1}. This conjecture was disproved by Katsman, Tsfasman and Vl\u{a}du\c{t} \cite{katsman-tsfasman-vladut-1992} using algebraic-geometric codes which beat the Gilbert-Varshamov bound. To the best of our knowledge, no other disproof is known in the literature. Theorem \ref{th:GV-lower} proves that $E(R)>\Eex^\infty(R)$ for the considered channel and hence it offers a second disproof of the conjecture, which only uses an elementary extension of the Gilbert-Varshamov procedure carefully  tuned for the particular case at hand.
\end{remark}

We finally comment the bound stated in Proposition \ref{prop:straight_line} showing how it is derived and why it is not trivial to improve it. 
For the particular channel considered, the sphere packing bound is essentially trivial; it states that $E(R)\leq 0$ above capacity, that is $R>\log(5/2)$, while $E(R)\leq \infty$ below capacity. On the other hand, Lov\'asz' proof of the zero-error capacity implies that $E(R)$ is finite for $R>C_0=\log(5)/2$. Finding good upper bounds on $E(R)$ in the range $\log(5)/2<R<\log(5/2)$ appears to be a difficult problem. To the best of our knowledge the most effective bound to date is obtained as follows. For $R>\log(5)/2$ at least two codewords are confusable and, from the point of view of these two codewords, the channel is like a binary erasure channel. Hence, in the extreme case when the two codewords are confusable but they differ in all positions, the probability of error is just the probability that all symbols are erased, that is $2^{-n}$. This implies that $E(R)\leq \log2$ for $R>\log(5)/2$. Using the straight line bound \cite{shannon-gallager-berlekamp-1967-1} we deduce the result of Proposition \ref{prop:straight_line}.

We observe that we have considered the optimistic condition where all confusable pairs of codewords differ in all possible positions. This may look too optimistic, but we point out that any sequence is confusable with $2^n$ other sequences which differ in all single position from the considered one. So, it is not obvious at all how we can improve the bound by reasoning along this line. We would need to prove that for $R>\log(5)/2$ there are sequences which are both confusable and differ only in a fraction $\delta<1$ of the positions. This is precisely what we will do using a linear programming approach in Section \ref{sec:converse-proof}.

\section{Proof of Theorem \ref{th:GV-lower}}
\label{sec:lower_bound}
We upper bound the error probability for a random code by using a standard union bound on the probability of confusion among single pairs of codewords. The code is built using a Gilbert-Varshamov-like procedure, though we exploit carefully the properties of the channel to add some structure to the random code and obtain better results than just picking random independent codewords. 

Consider a code with $M$ codewords $\bm{x}_1,\ldots,\bm{x}_M$. 
Let us first consider the probability $P(\bm{x}_j|\bm{x}_i)$ that a codeword $\bm{x}_i \in\mathbb{Z}_5^n$ sent through the channel is incorrectly decoded as a second codeword $\bm{x}_j \in\mathbb{Z}_5^n$. This is possible only if the two codewords are confusable, which means that their coordinates are all pairwise confusable. As for the problem of discriminating among these two codewords, the channel is equivalent to an erasure channel with erasure probability $1/2$. So, the sequence $\bm{x}_i$ can be incorrectly decoded as sequence $\bm{x}_j$ only if all differences are erased, which happens with probability $2^{-\dH(\bm{x}_i,\bm{x}_j)}$. So, we have 
\begin{equation*}
P(\bm{x}_j|\bm{x}_i)\leq
\begin{cases}
2^{-\dH(\bm{x}_i,\bm{x}_j)} & \bm{x}_i,\bm{x}_j\mbox{ confusable}\\
0 & \bm{x}_i,\bm{x}_j\mbox{ not confusable}
 \end{cases}
\end{equation*}
where $\dH$ is the Hamming distance. We can rewrite this in a simpler form if we introduce a notion of distance $d:\mathbb{Z}_5\times \mathbb{Z}_5\to\{0, 1,\infty\}$
\begin{equation*}
d(x_1,x_2)=
\begin{cases}
0 & x_1=x_2\\
1 & x_1-x_2=\pm 1\\
\infty & x_1-x_2\neq \pm 1
 \end{cases}
\end{equation*}
and then extend it additively to sequences in $\mathbb{Z}_5^n$
\begin{equation*}
d(\bm{x}_1,\bm{x}_2)=\sum_k d(x_{1,k},x_{2,k}).
\end{equation*}
Using this definition we can rewrite
\begin{equation*}
P(\bm{x}_j|\bm{x}_i)\leq
2^{-d(\bm{x}_i,\bm{x}_j)}.
\end{equation*}
The average probability of error can then be bounded using the union bound as
\begin{align*}
\Pe & \leq \frac{1}{M}\sum_{i\neq j} 2^{-d(\bm{x}_i,\bm{x}_j)}\\
& =  \sum_{z=0}^n A_z 2^{-z}\,,
\end{align*}
where $A_z$ is the spectrum of the code
\begin{equation*}
A_z=\frac{1}{M}\left|\{(i,j):  d(\bm{x}_i,\bm{x}_j)=z\} \right|.
\end{equation*}

Consider now linear codes of length $n'=2 n$ with $(n+k)\times 2n$ generator matrix of the form
\begin{equation*}
G^+=\left( 
\begin{array}{cc}
I_{n} &  2I_{n}\\
0 & G
\end{array}\right)
\end{equation*}
where $I_n$ is the $n\times n$ identity matrix and $G$ is a $k\times n$ matrix. We will study the family of random codes obtained when $G$ is a  random matrix with uniform independent entries in $\mathbb{Z}_5$. Note that this corresponds to taking $5^k$ randomly shifted versions of the $n$-fold cartesian power of Shannon's zero-error code of length 2 \cite{shannon-1956}.
Since we focus on linear codes, the $A_z$'s take the simpler form $A_z = \left|\{i :  w(\bm{x}_i)=z\} \right|$,
where we set $w(\bm{x}_i)=d(\bm{x}_i,\bm{0})$, the weight of the codeword (and similarly $w(x)=d(x,0)$).

We now proceed to the study of $A_t$.
We can decompose any information sequence $\bm{u}\in \mathbb{Z}_5^{n+k}$ in two parts, $\bm{u}=(\bm{u}_1, \bm{u}_2)$, with $\bm{u}_1 \in \mathbb{Z}_5^n$ and $\bm{u}_2\in\mathbb{Z}_5^k$. The associated codeword $\bm{v}=\bm{u}G^+$ can be correspondingly decomposed in two parts $\bm{v}=(\bm{v}_1, \bm{v}_2)$ with $\bm{v}_1=\bm{u}_1$ and $\bm{v}_2=2\bm{u}_1+\bm{u}_2G$. Call $\bm{\nu}=\bm{u}_2G$. We now relate the weight $w(\bm{v})$ to the \emph{Hamming weight} $\wH(\bm{\nu})$ and to the form of $\bm{u}_1$.
Note in particular that we can write
\begin{equation*}
w(\bm{v})=\sum_{i=1}^n w((v_{1,i},v_{2,i}))
\end{equation*}
and that
\begin{equation*}
(v_{1,i},v_{2,i})=u_{1,i}(1,2)+(0,\nu_i)\,.
\end{equation*}

Note first that $w(\bm{v})=\infty$ if $u_{1,i}=\pm 2$ for some $i$. So, for the study of $A_z$ we need only consider the cases $u_{1,i}\in \{0,\pm 1\}$. Consider first the case when $\nu_{i}=0$. If $u_{1,i}=0$ then $w(v_{1,i})=w(v_{2,i})=0$ while if $u_{1,i}=\pm 1$ then $w(v_{2,i})$ is infinite. So, if $\nu_{i}=0$ one choice of $u_{1,i}$ gives no contribution to $w(\bm{v})$ while all other choices lead to $w(\bm{v})=\infty$, and hence give no contribution to $A_z$ for any finite $z$.
Consider then the case of a component $\nu_{i}\neq 0$. It is not too difficult to check that one choice of $u_{1,i}$ in $\{0,\pm 1\}$ gives $w(v_{1,i})=w(v_{2,i})=1$, one gives $w(v_{1,i})=1$ and $w(v_{2,i})=0$ or vice-versa, and the remaining one gives  $w(v_{2,i})=\infty$.
So, if $\nu_{i}\neq 0$ one choice of $u_{1,i}$ contributes $1$ to $w(\bm{v})$, one choice of $u_{1,i}$ contributes $2$, while all other choices lead to $w(\bm{v})=\infty$, and hence give no contribution to $A_z$ for any finite $z$.

So, for a fixed $\bm{\nu}$ of Hamming weight $d$, and for a fixed $t\in\{1,2,\ldots,d\}$, there are $\binom{d}{t}$ vectors $\bm{u}_1$ which give codewords $\bm{v}$ of weight $2t+(d-t)=d+t$. If $B_d$ is the number of sequences $\bm{u}_2$ which lead to a $\bm{\nu}$ of Hamming weight $d$, then we have
\begin{equation}
\Pe\leq \sum_{d=1}^n \sum_{t=1}^d B_d \binom{d}{t} 2^{-(d+t)}\,.\label{eq:SumErLB}
\end{equation}
But $B_d$ is now simply the spectrum of the linear code with generator matrix $G$, and it is known from the Gilbert-Varshamov procedure that as we let $n$ and $k$ grow to infinity with ratio $k/n\to r$, matrices $G$ exist for which
\begin{equation*}
B_{\delta n}=
\begin{cases}
0 & \mbox{if } \delta < \delta_{GV}(r)\\
5^{n(r-1)}\binom{n}{\delta n}4^{n\delta}(1+o(1)) & \mbox{if } \delta\geq  \delta_{GV}(r)
\end{cases}
\end{equation*}
where $\delta_{GV}(r)$ is the Gilbert-Varshamov bound at rate $r$ determined implicitly by the relation
\begin{equation*}
r\log(5)=\log(5)-H_2(\delta_{GV}(r))+2\delta_{GV}(r)\,.
\end{equation*}
Defining $\delta=d/n$, $\tau=t/d$ and $r=k/n$, the probability of error is bounded to the first order in the exponent by the largest term in the sum \eqref{eq:SumErLB} as
\begin{multline*}
\frac{1}{n}\log \Pe  \leq \max_{\delta\geq \delta_{GV}(r), \tau\in[0,1]} [\log(5)(r-1)+H_2(\delta)+2\delta \\+\delta H_2(\tau)-(\delta+\delta\cdot\tau)] + o(1)\,.
\end{multline*}
The maximum over $\tau$ is obtained by maximizing $H_2(\tau)-\tau$, which gives $\tau=1/3$, independently of $\delta$. So, we are left with the maximization
\begin{equation*}
\max_{\delta\geq \delta_{GV}(r)} [\log(5)(r-1)+H_2(\delta)+\delta(h(1/3)+2/3)]\,.
\end{equation*}
The argument is increasing for $\delta\leq 3/4$, where it achieves the maximum value  $\log(5)(r-1)+2$, and decreasing for $\delta>3/4$. So, the maximizing $\delta$ is $\delta=3/4$ if $3/4\geq\delta_{GV}(r)$ and $\delta_{GV}(r)$ otherwise.
Combining these facts, after some computation we find
\begin{equation*}
\frac{1}{n}\log\Pe\leq
\begin{cases}
(\log(5)(r-1)+2) +o(1),& \delta_{GV}(r)\leq 3/4\\
\delta_{GV}(r)(H_2(1/3)-4/3) +o(1),&  \delta_{GV}(r)> 3/4 \,.
\end{cases}
\end{equation*}
Considering that the block length is $2n$ and the rate of the global code is $R=\log(5)(1+r)/2$, after some simple algebraic manipulations we obtain the claimed bound.

\section{Proof of Theorem \ref{th:ELP}}
\label{sec:converse-proof}

The upper bound we derive here is based on bounding the maximum probability of error over all codewords $\Pemax$, which in turn we bound in terms of the minimum distance of codes for the distance measure introduced in the previous section. Note in particular that we have 
\begin{equation*}
\Pemax\geq \max_{i\neq j}\frac{1}{2} \cdot 2^{-d(\bm{x}_i,\bm{x}_j)}\,.
\end{equation*}
Indeed, if there is no pair of confusable codewords, then the inequality is trivial, while if codewords $i$ and $j$ are confusable, any (possibly randomized) decoder will decode in error with probability at least $1/2$ either when codeword $i$ or codeword $j$ is sent. So, we can bound the reliability as
\begin{equation}
E(R)\leq \min_{i\neq j} \frac{1}{n}d(\bm{x}_i,\bm{x}_j) (1+o(1))\,.
\label{eq:Etod}
\end{equation}
The rest of this section is devoted to bounding the minimum distance. In particular we prove that codes for which 
\begin{equation*}
 \min_{i\neq j} \frac{1}{n}d(\bm{x}_i,\bm{x}_j) \geq \delta
\end{equation*}
have rate $R$ upper bounded as
\begin{equation}
R\leq  {1\over 2} \log 5 + R_{\text{LP1}}(\sqrt{5}, \delta)(1+o(1))\,.
\label{eq:R_LP'_RLP}
\end{equation}
Note that Theorem \ref{th:ELP} follows from equations \eqref{eq:Etod}-\eqref{eq:R_LP'_RLP}. 

Our bound is based on $\theta$ functions and Delsarte's linear programming bound \cite{delsarte-1973}, but it is easier to describe it in terms of Fourier transforms. We set here $q=5$ and give a proof in terms of $q$ which also holds for any other odd larger value.

For any 
$f:\mathbb{Z}_q^n\to \mathbb{C}$ we define its Fourier transform as
\[
\hat f(\bm{\omega}) = \sum_{\bm{x}\in \mathbb{Z}_q^n} f(\bm{x})e^{\frac{2\pi i}{q}<\bm{\omega},\bm{x}>},\quad \bm{\omega}\in \mathbb{Z}_q^n
\]
where the non-degenerate $\mathbb{Z}_q$-valued bilinear form is defined as usual
$$<\bm{x},\bm{y}>\eqdef \sum_{k=1}^n x_k y_k\,. $$ 
We also define the inner product as follows
$$ (f,g) \eqdef q^{-n} \sum_{\bm{x}\in \mathbb{Z}_q^n} \bar f(\bm{x}) g(\bm{x})\,. $$

The starting point is a known rephrasing of linear programming bound. Let $\matc$ be a code with minimum distance at
least $d$. Let $f$ be such that $f(\bm{x})\leq 0$ if $d(\bm{x},\bm{0})=w(\bm{x})\geq d$, $\hat{f}\geq 0$ and $\hat{f}(\bm{0})>0$. Then,  consider the Plancherel  identity
\begin{equation*}
(f* 1_\matc, 1_\matc)=q^{-n}(\hat{f}\cdot\widehat{1_\matc}, \widehat{1_\matc})\,,
\end{equation*}
where $1_A$ is the indicator function of a set $A$. Upper bounding the left hand side by $|\matc|f(\bm{0})$ and lower
bounding the right hand side by the zero-frequency term $q^{-n}\hat{f}(\bm{0})|\matc|^2$, one gets
\begin{align}\label{eq:f0fhat0}
%|\matc| & \leq \min \left\{q^n \frac{f(\bm{0})}{\hat{f}(\bm{0})}: %f(\bm{x}) \le 0\mbox{~for~} w(\bm{x})\geq d, \hat f \ge 0, \hat{f}%(\bm{0})>0 \right\}\,.
|\matc| & \leq \min q^n \frac{f(\bm{0})}{\hat{f}(\bm{0})}.
\end{align}
The proof of our theorem is based on a choice $f$ which combines Lov\'asz' assignment used to obtain his bound on the zero-error capacity with the one used in \cite{mceliece-et-al-1977} to obtain bounds on the minimum distance of codes in Hamming spaces.

Observe first that Lov\'asz assignment can be written in one dimension ($n=1$) as
\[
g_1(x) = 1_0(x)+ \varphi  1_{\pm 1} (x),\quad x\in \mathbb{Z}_q\,,
\]
where  $\varphi=(2\cos(\pi/q))^{-1}$. This gives 
\begin{equation*}
\widehat{g_1}(\omega)=1+2\varphi \cos(2 \pi \omega/q), \quad \omega\in \mathbb{Z}_q.
\end{equation*}
Correspondingly, define the $n$-dimensional assignment 
\[
g(\bm{x}) = \prod_{j=1}^n g_1(x_j), \quad 
\hat g(\bm{\omega}) = \prod_{j=1}^n\widehat{g_1}(\omega_j), \quad \bm{x},\bm{\omega} \in \mathbb{Z}_q^n.
\]
Note that $\widehat{g_1}\geq 0$ and, additionally, $\widehat{g_1}(\omega)=0$ for $\omega=\pm c$, with $c=(q-1)/2$. So, $\hat{g}\geq 0$, with $g(\bm{\omega})=0$ if $\bm{\omega}$ contains any $\pm c$ entry.
Since $g(\bm{x})=0$ for $\bm{x}\notin \{0,\pm 1\}^n$, $g$ satisfies all the properties required for $f$ in the case  $d=\infty$, and when used in place of $f$ in \eqref{eq:f0fhat0} it gives Lov\'asz' bound
\begin{align*}
|\matc| & \leq q^n\frac{g(\bm{0})}{\hat{g}(\bm{0})}\\
& = q^n \left(\frac{\cos(\pi/q)}{1+\cos(\pi/q)}\right)^n
\end{align*}
for codes of infinite minimum distance.

For the case of finite $d\leq n$, we build a function $f$ of the form $f(\bm{x})=g(\bm{x})h(\bm{x})$, for an appropriate $h(\bm{x})$. In
particular, since $g(\bm{x})$ is non-negative and already takes care of setting $f(\bm{x})$ to zero if $x\notin \{0,\pm1\}^n$, it suffices to choose $h$ such that $h(\bm{x})\leq 0 $ whenever $\bm{x}\in\{0,\pm1\}^n$ contains at least $d$ entries with value $\pm1$. We restrict attention to $h$ such that $\hat{h}\geq 0$, so that $\hat{f}=q^{-n}\hat{g}*\hat{h}\geq 0$. In particular, we consider functions $h$ whose Fourier transform is constant on each of the following ``spheres'' in $\mathbb{Z}_q^n$ 
%composed of $\ell$ values $\pm c$ and $n-\ell$ zero values:
\[
S_\ell^c = \{\bm{\omega}:|\{i:\omega_i=\pm c \}|=\ell,\;  |\{i:\omega_i=0 \}|=n-\ell\}\,,\quad \ell=0,\ldots, n\,,
\]
and zero outside. This choice is motivated by the fact, observed before, that $\hat g_1(\pm c)=0$. Restricting $\hat{h}$ to be null out of these spheres simplifies the problem considerably.
We thus define
\begin{equation}\label{eq:formofh}
\hat{h}(\bm{\omega})=\sum_{\ell=0}^n \hat{h}_\ell 1_{S_{\ell}^c}(\bm{\omega})\,,\quad h(\bm{x})=q^{-n}\sum_{\ell=0}^n \hat{h}_\ell \widehat {1_{S_{\ell}^c}}(\bm{x})\,,
\end{equation}
where $\hat{h}_\ell\geq 0$ and $\hat{h}_0>0$ will be optimized later.
Since $\hat{g}(\bm{\omega})=0$, $\bm{\omega}\in S_\ell\,, \ell>0$, setting $f(\bm{x})=g(\bm{x})h(\bm{x})$ gives
$\hat{f}(\bm{0})=q^{-n}(\hat{g}*\hat{h})(\bm{0})=q^{-n}\hat{g}(\bm{0})\hat{h}_0$. So, the bound \eqref{eq:f0fhat0} becomes 
\begin{equation*}
|\matc|\leq \left(q^n\frac{g(\bm{0})}{\hat{g}(\bm{0})}\right)\left(q^n \frac{h(\bm{0})}{\hat{h}_0}\right)\,.
\end{equation*}
The first term above is precisely Lov\'asz bound and corresponds, for $q=5$, to the $\frac{1}{2}\log(5)$ term in the right hand side of  \eqref{eq:R_LP'_RLP}. We now show that the second term corresponds to the linear programming bound of an imaginary ``Hamming scheme'' with a special non-integer alphabet size $q'=1+\cos(\pi/q)^{-1}$. To do this, define analogously to $S_\ell^c$ the spheres
 \[
S_u^1 = \{\bm{x}:|\{i:x_i=\pm 1 \}|=u,\;  |\{i:x_i=0 \}|=n-u\} \,.
\]
Our constraint is that $h(\bm{x})\leq 0$ if $\bm{x}\in S_u^1$, $u\geq d$. Direct computation shows that for  $\bm{x}\in S_u^1$,
\begin{align*}
\widehat {1_{S_{\ell}^c}}(\bm{x}) & =\sum_{j=0}^\ell \binom{u}{j}\binom{n-u}{\ell-j}(-1)^j2^\ell (\cos(\pi/q))^j\,,\qquad (\bm{x}\in S_u^1)\\
& = (2\cos(\pi/q))^\ell  K_\ell(u;q'),\qquad (q'=1+\cos(\pi/q)^{-1}\,),
\end{align*}
where $K_\ell(u;q')$ is a Krawtchouck polynomial of degree $\ell$ and parameter $q'$ in the variable $u$. We can thus define
\begin{equation*}
\Lambda(u)=h(\bm{x})\,, \bm{x}\in S_u^1\,,\qquad \lambda_\ell=q^{-n} (2\cos(\pi/q))^\ell\cdot  \hat{h}_\ell\,,
\end{equation*} 
and write
\begin{equation}
\label{eq:htoLPq'}
q^n\frac{h(\bm{0})}{\hat{h}_0} = \frac{\Lambda(0)}{\lambda_0}\,,
\end{equation}
where the conditions on $h$ can be restated as
\begin{align*}
\Lambda(u) & =\sum_{\ell=0}^n \lambda_\ell K_\ell(u;q')\,,\quad u=0,\ldots,n\,,\\
\lambda_\ell & \geq 0 \,,\quad \ell \geq 0\,,\\
\Lambda(u) & \leq 0\,,\quad u\geq d\,.
\end{align*}
So, the minimization of \eqref{eq:htoLPq'} is reduced to the standard linear programming problem for the Hamming space, though with a non-integer parameter $q'$. Since the construction of the polynomial used in \cite{mceliece-et-al-1977} and \cite{aaltonen1990new} can be applied verbatim for non-integer values of $q'$ (see also \cite{ismail-simeonov-1998} for the position of the roots of $K_\ell(u;q')$), the claimed bound follows.

%%%%%%%%%%%%%

%
%% References:
%% We recommend the usage of BibTeX:
%%
%\bibliographystyle{IEEEtran}
%\bibliography{bibeit}

\end{document}